\documentclass[10pt,twocolumn, nofootinbib]{revtex4-2}

\usepackage{assumptionsofphysics}
\usepackage{graphicx}
\usepackage{hyperref}
\hypersetup{
	colorlinks=true,
	citecolor=blue,
	urlcolor=blue,
	linkcolor=blue
}
\newcommand\partitle[1]{\textsc{#1}.}

\begin{document}

\title{Experimental verifiability and topology }
\author{Gabriele Carcassi, Christine A. Aidala}
\affiliation{Physics Department, University of Michigan, Ann Arbor, MI 48109}

\date{\today}

\begin{abstract}
We briefly show how the use of topological spaces and $\sigma$-algebras in physics can be rederived and understood as the fundamental requirement of experimental verifiability. We will see that a set of experimentally distinguishable objects will necessarily be endowed with a topology that is Kolmogorov (i.e. $T_0$) and second countable, which both puts constraints on well-formed scientific theories and allows us to give concrete physical meaning to the mathematical constructs. These insights can be taken as a first step in a general mathematical theory for experimental science.
\end{abstract}

\maketitle

\section{Introduction}

The overall structure (see \cite{aop-book} for more details) can be summed up in the following diagram that can be used as a guide throughout this note.

\begin{figure}[h]
	\includegraphics[width=\columnwidth]{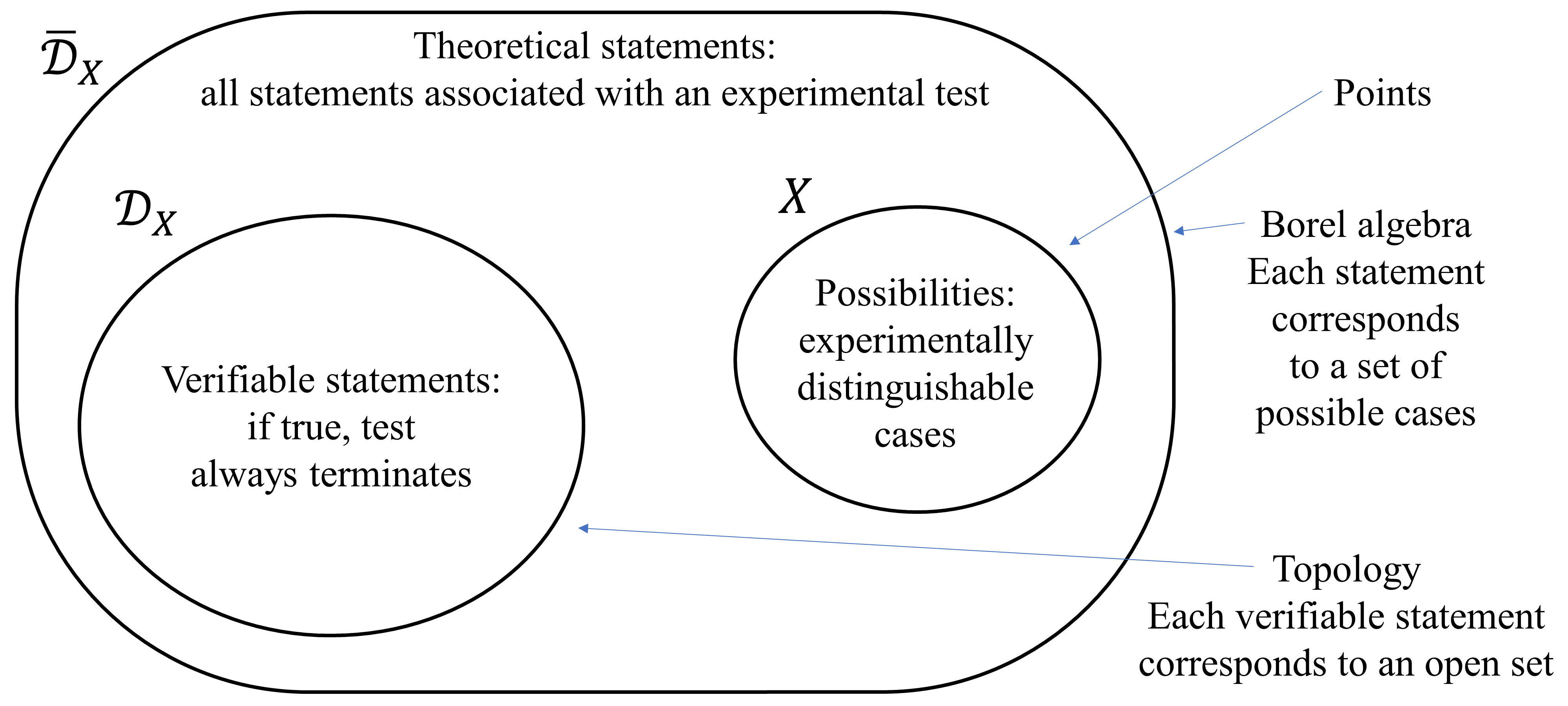}
\end{figure}

A theoretical domain $\tdomain_X$ contains all the statements about a scientific object that can be associated to an experimental test, which we call theoretical statements.\footnote{For example, all the possible statements about the mass of a particle.} Within this set we find the possibilities which describe the possible cases that can be experimentally distinguished\footnote{For example, \statement{the mass of the particle is exactly zero.}}. Not all tests are guaranteed to always terminate,\footnote{For example, infinite precision measurements are only possible as the limit of an infinite sequence of finite precision measurements.} so we find the subset $\edomain_X \subseteq \tdomain_X$ consisting of the verifiable statements, called experimental domain, for which successful termination coincides with the truth of the statement.\footnote{For example, \statement{the mass of the particle is less than 100 GeV.}}

The possibilities correspond to the points of our spaces. The experimental domain corresponds to the topology, where each open set corresponds to a verifiable statement.\footnote{For example, \statement{the mass of the particle is less than 100 GeV} corresponds to the open interval $(0, 100)$.} The theoretical domain corresponds to the Borel algebra.

\section{Logical context and fundamental axioms}

\partitle{Principle of scientific objectivity} The guiding principle is that \emph{science is universal, non-contradictory and evidence based.} This means that any scientific theory must specify a set of \textbf{statements} that are logically consistent, whose truth is the same for everybody and connected to experimental verification. This starting point is codified in the following axioms, which form the formal foundation of our theory.

\partitle{Axioms of logic} The first set of axioms ensure the theory is universal and non-contradictory. The \emph{axiom of context} tells us that statements are organized into a \textbf{logical context} $\logCtx$, and a well defined truth value for each statement must exist. This recognizes that the logical consistency and meaning of each statement relies on definitions and relationships that depend on the context\footnote{For example, if one is performing particle identification, the mass of an electron is assumed to be given. If one is measuring the mass, the identity is assumed to be given, while the value of the mass is unknown.}, on a set of statements.

The \emph{axiom of possibility} tell us that the statements in a context cannot be assigned arbitrary truth values, but only those consistent with the meaning defined by the context.\footnote{For example, \statement{the mass of the object is one kilogram} and \statement{the mass of the object is two kilograms} cannot both be true.} Therefore each context comes with a set $\pAss$ of possible assignments, where each possible assignment is a function $a : \logCtx \to \Bool$ that returns the truth value for each statement.

The \emph{axiom of closure} tells us that we can always find a statement whose truth arbitrarily depends on the truth of other statements.

\partitle{Logical relationships} The axioms of logic entail that each logical context is a complete Boolean algebra. They allow us to define useful relationships such as: equivalence $\equiv$ (i.e. statements that must be equally true or false); narrowness (i.e. a statement is more specific than another)\footnote{For example, \statement{the mass of the object is between 4 and 5 kilograms} is narrower than \statement{the mass of the object is between 0 and 100 kilograms}. Narrowness can also be understood as implication.}; compatibility $\comp$ (i.e. whether two statement can be true at the same time); independence $\indep$ (i.e. whether the truth of some statements leaves the truth of others unconstrained). This basic structure constrains any further mathematical structure as the structures will have to interact appropriately.

\partitle{Axioms of verifiability} This set of axioms capture the rules of experimental verifiability. The \emph{axiom of verifiability} tells us that some statements are \textbf{verifiable}, meaning we have a test at our disposal that will terminate successfully in finite time if and only if the statement is true. Note that the test is not guaranteed to terminate, to give an answer, if the statement is false.\footnote{For example, a test for \statement{there exists extra-terrestrial life} may terminate successfully if life is found, but it will not be conclusive if life is not found. On the other hand, \statement{the mass of the particle is exactly 0 eV} can be experimentally disproved, but can never be confirmed due to finite precision.} Therefore the negation of a verifiable statement is not necessarily verifiable. It also tells us that a statement equivalent to a verifiable statement is verifiable, as we can use the test of the second to test the first.

The remaining two axioms state that the finite conjunction and the countable disjunction of verifiable statements are verifiable. That is, we can use the tests of the given statements to construct the test for the logical AND (i.e. check that all tests terminate successfully) and the logical OR (i.e. check that at least one test terminates successfully). The disjunction can be extended to a countable set because we only need one test to terminate successfully in a finite time, not all of them.

\partitle{Decidability} A \textbf{falsifiable} statement is one whose negation is verifiable. A \textbf{decidable} statement is one that is both verifiable and falsifiable. The finite conjunction/disjunction and negation of decidable statements are decidable.

\partitle{Different algebras} The logical context as a whole, the set of verifiable statements and the set of decidable statements follow different algebras according to Table \ref{algebras}.

\begin{table}[h]
	\centering
	\begin{tabular}{p{0.10\textwidth} p{0.05\textwidth} p{0.10\textwidth} p{0.10\textwidth} p{0.10\textwidth}}
		Operator & Gate & Statement & Verifiable Statement & Decidable Statement  \\ 
		\hline 
		Negation & NOT & allowed & disallowed & allowed \\ 
		Conjunction & AND & arbitrary  & finite & finite \\ 
		Disjunction & OR & arbitrary  & countable & finite \\ 
	\end{tabular}
	\caption{Comparing algebras of statements.}
	\label{algebras}
\end{table}

\section{Experimental and theoretical domains}

\partitle{Experimental domains} The next step is to group together verifiable statements that characterize the same physical object (e.g. all the statements that describe the state of a system or the properties of an object). We call such a group an \textbf{experimental domain} $\edomain_X \subseteq \logCtx$ and it will have at least two characteristics. First of all, it will be closed under finite conjunction (i.e. logical AND) and countable disjunction (i.e. logical OR) since these are exactly the operations that give us other verifiable statements. Second, it cannot be too big: we must be able to explore it fully given an arbitrarily long amount of time. Formally, it can be generated from a countable subset of verifiable statements through finite conjunction and countable disjunction, that is, it must have a countable \textbf{base}. This means the tests for all statements in the domain can be generated from the tests of the base, therefore running those countably many tests is enough to test the whole domain.

\partitle{Theoretical domains} While the verifiable statements define all the testing procedures that are available in some domain, they are not the only valid statements in a scientific theory.\footnote{For example, \statement{the mass of the photon is exactly 0 eV} is not something we can verify due to the limited precision of our statements, yet it is something we want to be able to say in our theory.} We are interested in all statements that can be assigned an experimental test, regardless of its terminating conditions. We call these \textbf{theoretical statements} and we group them into a \textbf{theoretical domain} $\tdomain_X$. This is generated from the experimental domain by closing under negation and countable disjunction.

\partitle{Possibilities} Within a theoretical domain we find the \textbf{possibilities} $X$ of the domain $\edomain_X$. These are statements that give the narrowest description possible and, if assumed to be true, tell us the truth value for all other statements. The set of possibilities represents all the possible cases that can be distinguished by the experimental domain.\footnote{For example, if the domain characterizes the state of a bead along a wire, ``the position of the bead is 3 meters and the velocity is zero meters per second'' is a possibility because it characterizes the truth value of all possible experimentally verifiable finite precision descriptions.}

\partitle{Correspondence to topologies and $\sigma$-algebras} Every statement in the domain can be characterized by the set of possibilities for which the statement is true. Therefore, we have a correspondence between sets and statements where the logic operations become the set operations. Under this correspondence, the experimental domain induces a topology where each open set represents a verifiable statement, while the theoretical domain induces a $\sigma$-algebra where each Borel set is a theoretical statement. The topology is at least $T_0$ and second countable while the Borel algebra is countably generated (and therefore separable), which makes them extremely well-behaved.

\partitle{On the significance of mathematical structures} \textbf{The existence of well-behaved topologies and $\sigma$-algebras is a direct consequence of the requirement of experimental verifiabiliy.}\footnote{Philosophically, one may say that it is not an ontological attribute of the system being studied, but rather a characterization of its epistemically accessible characteristics. There is no reason to assume that ontological entities would follow such constraint.} Another important aspect is that \textbf{the points are not the starting point, but rather they are generated from the verifiable statements.} If we change what is being characterized (e.g. add measurements of spin to those of the position and momentum of an electron) the points of the space will change.\footnote{Philosophically, the points themselves cannot be understood as some complete ontological entity, but rather the epistemically accessible characteristics under a set of given circumstances.}

\partitle{Unphysical statements} Our construction makes it clear that non-Borel sets play no role in physical theories, not because they are hard to work with, but because they are unphysical. Note that the theoretical domain is closed under countable operations, while the context is closed under arbitrary operations. This means that, in line of principle, there may be statements that depend on the experimental ones that are not theoretical.  In fact, if the possibilities are uncountable (e.g. we use real numbers to identify them, as we do when using position for states) such statements are guaranteed to exist. Since they are not theoretical, they are not associated to a testing procedure: they are unphysical and are automatically excluded by our construction. It is a feature of our theory that tells us exactly which mathematical objects are physically meaningful and which aren't.

\partitle{Maximum cardinality for the possibilities} Since the domain is generated by a countable set of verifiable statements, the possibilities must be distinguished with a countable sequence of Boolean values. This means that the cardinality of the possibilities cannot exceed that of the continuum (an infinite sequence of zeros and ones is a real number expressed in base 2), therefore infinities of higher order and all associated problems do not play a role in science.

\partitle{Verifiable, falsifiable and undecidable parts} Given each theoretical statement, we can define its verifiable part (i.e. the set of possibilities for which its test terminates successfully), its falsifiable part (i.e. the test terminates successfully) and the undecidable part (i.e. the test does not terminate). Topologically, these correspond to the interior, exterior and boundary of the corresponding Borel set.\footnote{Note there are domains that contain undecidable statements, those for which the test never terminates. For example, \statement{the mass of the proton is rational in units of eV} is undecidable: any finite precision measurement will be consistent with infinitely many rational and irrational possibilities. Mathematically, the boundary of the rationals is all the reals.}

\partitle{Discreteness is decidability} A decidable statement will correspond to a clopen set. A topology is discrete if all sets are clopen, which means all the statements are decidable.\footnote{Many physicists confuse discreteness with countability, and characterize the rationals as discrete. This depends on how the rationals are measured. If the numerator and denominator are measured independently, then an infinite precision measurement is made on the rationals: the topology is discrete. If the value is measured with bounds, no infinite precision measurement is made and the value can be excluded but never fully verified: the topology is the one inherited from the reals.} It can be shown that a decidable domain, one where all statements are decidable, can only have up to countably many possibilities. Therefore, if the possibilities are uncountable, not all statements can be decidable.

\section{Experimental relationships and domain combination}

\partitle{Inference relationships} Intuitively, an \textbf{ inference relationship} between two domains $\edomain_X$ and $\edomain_Y$ allows us to infer truths on the first by performing experimental verification on the second. That is, an inference relationship is a map $\erel : \edomain_Y \to \edomain_X$ that given a verifiable statement $\stmt_Y \in \edomain_Y$ gives us an equivalent statement $\stmt_Y \equiv \erel(\stmt_Y) \in \edomain_X$.\footnote{For example, by looking at the height of a column of mercury we are able to infer the temperature of the body.}

\partitle{Causal relationship} Intuitively, a \textbf{causal relationship} between two domains $\edomain_X$ and $\edomain_Y$ means that if we determine which possibility $x \in X$ is true in the first domain, then this will also tell us what possibility $y \in Y$ is true in the second. That is, a causal relationship is a map $f : X \to Y$ such that $x \narrower f(x)$: if $x$ is true so is $f(x)$. An important result is that \textbf{causal relationships must be continuous} in the natural topology. Continuity, therefore, is a necessary requirement, not a desired property. We stress that this is topological continuity, and not necessarily analytical continuity.\footnote{Analytical continuity breaks down on discrete points, which means on decidable possibilities. For example, whether a mixture of water, ice and water vapor is at the triple point of water is decidable, and therefore we can expect analytical discontinuities there, and in other phase transitions.}

\partitle{Experimental relationships and domain dependence} An important result is that \textbf{two domains admit an inference relationship if and only if they admit a causal relationship}. We use the term \textbf{experimental relationship} to describe the link between the domains and we say $\edomain_X$ \textbf{depends on} $\edomain_Y$, noted $\edomain_X \subseteq \edomain_Y$. Note that the inference works in the opposite direction of the cause.

The link between inference and causation is very important. Inference relationships are defined over verifiable statements, which are our link to experimental verification, and therefore have a higher status in the theory. Unfortunately they are highly redundant\footnote{For example, an inference relationship will tell us how every finite precision measurement of the height of a mercury column would correspond to a finite precision measurement of temperature. As the finite intervals overlap in infinitely many ways, the overlaps are specified again and again.} and therefore harder to work with. On the other hand, causal relationships are defined over the possibilities, which are less directly related to experimental verification, but they are a lot more compact\footnote{For example, a causal relationship will tell us what infinitely precise height of the mercury column will correspond to each infinitely precise value of temperature. Each value is mapped once to another value.} and therefore more practical mathematically. The link between them tells us that we can always go between one and the other, so we can use causal relationships knowing that they are as experimentally well-defined as their counterparts.

\partitle{Domain combination} Given a countable set of experimental domains $\{\edomain_X\}_{i=1}^{\infty}$, the combined domain is the experimental domain of all the verifiable statements generated by them. Formally, it is the closure under finite conjunction and countable disjunction.\footnote{For example, we take the domains of different quantities of an object to form the domain of its state.} The possibilities of the combined domain can be understood as a subset of the Cartesian product of the possibilities of all domains.\footnote{For example, the value of all the quantities provided by each domain.} The fewer correlations between the different domains, the bigger the set of combined possibilities. If the domains are independent, all combinations in the Cartesian product will correspond to a possibility for the combined domain. Moreover, the topology will be the product topology of the domains. If one domain depends on another, their combined domain will be equivalent to the second: no new possibilities are added. If the domains are incompatible, then the possibilities of the combined domain are the disjoint union and the topology will be the disjoint union topology.

\partitle{Relationship domain} Given two experimental domains, we can construct a domain where the possibilities are the possible relationships between the domains. We call this the \textbf{relationship domain}. We can proceed recursively and construct domains of higher order functions, without going outside the mathematical framework. There is, however, a caveat: the construction of this domain requires the existence and construction of a logical meta-context. Therefore, while we can show that mathematically the construction can always be done, there is no guarantee that the statements are verifiable. That is, we cannot construct tests that verify relationships between quantities from tests that simply verify the values of the quantities.\footnote{That is, being able to verify particular values of temperature and pressure is not enough to be able to verify whether to a particular value of temperature always corresponds a particular value of pressure.}

\bibliography{bibliography}

\begin{thebibliography}{1}%
\makeatletter
\providecommand \@ifxundefined [1]{%
 \@ifx{#1\undefined}
}%
\providecommand \@ifnum [1]{%
 \ifnum #1\expandafter \@firstoftwo
 \else \expandafter \@secondoftwo
 \fi
}%
\providecommand \@ifx [1]{%
 \ifx #1\expandafter \@firstoftwo
 \else \expandafter \@secondoftwo
 \fi
}%
\providecommand \natexlab [1]{#1}%
\providecommand \enquote  [1]{``#1''}%
\providecommand \bibnamefont  [1]{#1}%
\providecommand \bibfnamefont [1]{#1}%
\providecommand \citenamefont [1]{#1}%
\providecommand \href@noop [0]{\@secondoftwo}%
\providecommand \href [0]{\begingroup \@sanitize@url \@href}%
\providecommand \@href[1]{\@@startlink{#1}\@@href}%
\providecommand \@@href[1]{\endgroup#1\@@endlink}%
\providecommand \@sanitize@url [0]{\catcode `\\12\catcode `\$12\catcode
  `\&12\catcode `\#12\catcode `\^12\catcode `\_12\catcode `\%12\relax}%
\providecommand \@@startlink[1]{}%
\providecommand \@@endlink[0]{}%
\providecommand \url  [0]{\begingroup\@sanitize@url \@url }%
\providecommand \@url [1]{\endgroup\@href {#1}{\urlprefix }}%
\providecommand \urlprefix  [0]{URL }%
\providecommand \Eprint [0]{\href }%
\providecommand \doibase [0]{https://doi.org/}%
\providecommand \selectlanguage [0]{\@gobble}%
\providecommand \bibinfo  [0]{\@secondoftwo}%
\providecommand \bibfield  [0]{\@secondoftwo}%
\providecommand \translation [1]{[#1]}%
\providecommand \BibitemOpen [0]{}%
\providecommand \bibitemStop [0]{}%
\providecommand \bibitemNoStop [0]{.\EOS\space}%
\providecommand \EOS [0]{\spacefactor3000\relax}%
\providecommand \BibitemShut  [1]{\csname bibitem#1\endcsname}%
\let\auto@bib@innerbib\@empty
\bibitem [{\citenamefont {Carcassi}\ and\ \citenamefont
  {Aidala}(2021)}]{aop-book}%
  \BibitemOpen
  \bibfield  {author} {\bibinfo {author} {\bibfnamefont {G.}~\bibnamefont
  {Carcassi}}\ and\ \bibinfo {author} {\bibfnamefont {C.~A.}\ \bibnamefont
  {Aidala}},\ }\href {https://doi.org/10.3998/mpub.12204707} {\emph {\bibinfo
  {title} {Assumptions of Physics}}}\ (\bibinfo  {publisher} {Michigan
  Publishing},\ \bibinfo {year} {2021})\BibitemShut {NoStop}%
\end{thebibliography}%

\end{document}